\documentclass[aip,graphicx]{revtex4-2}

\usepackage{textcomp}
\usepackage{gensymb}
\usepackage{graphicx}
\usepackage[T1]{fontenc}

\begin{document}

 \author{Kartik Gaur}
 \author{Ching-Wen Shih}
 \author{Imad Limame}
 \author{Aris Koulas-Simos}
 \author{Niels Heermeier}
 \author{Chirag C. Palekar}
 \author{Sarthak Tripathi}
 \author{Sven Rodt}
 \author{Stephan Reitzenstein}
 \email[Corresponding author: ]{stephan.reitzenstein@physik.tu-berlin.de}
\affiliation{Institut für Festkörperphysik, Technische Universität Berlin, Hardenbergstraße 36, D-10623 Berlin, Germany
}%

\title{High-$\beta$ lasing in photonic-defect semiconductor-dielectric hybrid microresonators with embedded InGaAs quantum dots} 

\date{\today}

\begin{abstract}
We report an easy-to-fabricate microcavity design to produce optically pumped high-$\beta$ quantum dot microlasers. Our cavity concept is based on a buried photonic-defect for tight lateral mode confinement in a quasi-planar microcavity system, which includes an upper dielectric distributed Bragg reflector (DBR) as a promising alternative to conventional III-V semiconductor DBRs. 
Through the integration of a photonic-defect, we achieve low mode volumes as low as 0.28 $\mu$m\textsuperscript{3}, leading to enhanced light-matter interaction, without the additional need for complex lateral nanoprocessing of micropillars. We fabricate semiconductor-dielectric hybrid microcavities, consisting of Al\textsubscript{0.9}Ga\textsubscript{0.1}As/GaAs bottom DBR with 33.5 mirror pairs, dielectric SiO\textsubscript{2}/SiN\textsubscript{x} top DBR with 5, 10, 15, and 19 mirror pairs, and photonic-defects with varying lateral size in the range of 1.5 $\mu$m to 2.5 $\mu$m incorporated into a one-$\lambda/n$ GaAs cavity with InGaAs quantum dots as active medium. The cavities show distinct emission features with a characteristic photonic defect size-dependent mode separation and \emph{Q}-factors up to 17000 for 19 upper mirror pairs in excellent agreement with numeric simulations. Comprehensive investigations further reveal lasing operation with a systematic increase (decrease) of the $\beta$-factor (threshold pump power) with the number of mirror pairs in the upper dielectric DBR. Notably, due to the quasi-planar device geometry, the microlasers show high temperature stability, evidenced by the absence of temperature-induced red-shift of emission energy and linewidth broadening typically observed for nano- and microlasers at high excitation powers.
\end{abstract}


\maketitle 

\section{INTRODUCTION}

Significant advances in optoelectronic and quantum nanophotonic devices have been enabled over the past few decades by the development of high-quality micro- and nanocavities with integrated quantum emitters~\cite{Chow2018}. Due to their small mode volumes and high cavity \emph{Q}-factors \cite{vahala2003optical, microcavities_book}, such structures are characterized by enhanced light-matter interaction in the weak and strong coupling regimes of quantum electrodynamics (cQED)~\cite{Khitrova2006}. While the coherent light-matter interaction observed in the strong-coupling regime of cQED is most interesting in the area of fundamental research \cite{Reitzenstein2012}, weak coupling and the associated Purcell effect are nowadays present in many resonator-based nanophotonic devices such as quantum light sources \cite{Senellart2017} and high $\beta$-microlasers \cite{StephanWileyReview2021}, leading to a higher source brightness and a lower lasing threshold, respectively.

Many of these cavity-enhanced devices use self-assembled semiconductor quantum dots (QDs) as optically active elements and sophisticated nanotechnology platforms are required for their fabrication. For example, deterministic nanoprocessing technologies are nowadays used to maximize the photon extraction efficiency of cavity-based quantum light sources (e.g. Ref.~\cite{Rodt_2020b} with references therein). Furthermore, special care has to be taken in the lateral structuring process using, for instance, reactive ion etching (RIE) to ensure smooth sidewalls for tight mode confinement with low optical losses in the case of micropillar cavities. 

Interestingly, little attention has been paid to quasi-planar microcavities with low mode volumes, namely photonic defect-based cavities, which do not require advanced lateral patterning for three-dimensional mode confinement. This interesting cavity concept was first proposed and implemented for exciton-polariton devices~\cite{ElDaifPolaritonBox} and later also for QD-based single-photon sources \cite{ding2013vertical} to improve their photon extraction efficiency. Realizations of this concept include cavity-enhanced single-photon sources based on unintentionally grown photonic defects \cite{Maier2014} and wet-chemically etched nano- and microlenses \cite{MichlerGaussian} to realize 3D mode confinement in a quasi-planar microcavity geometry. Here, by varying the geometry of these photonic defects, one can flexibly tailor the spatial distribution of the 3D confined-cavity modes \cite{ding2013vertical}. Hence, by meticulously designing the defect shape, it is possible to control and optimize optical mode confinement in the active region where QDs reside. This can help to achieve very low mode volumes on a scale of the cube-wavelength in the quasi-planar device geometry while maintaining high \emph{Q}-factors. Consequently, one obtains improved light-matter interaction in the cQED regime, which, for example, increases the $\beta$-factor of microlasers. The defect shape can also influence the photon extraction efficiency, thus enabling efficient light extraction from the device, which is of particular interest for developing bright quantum light sources. The confinement of light within defect cavities with Gaussian far-field emission patterns can allow for accurate mode-matching with optical fibers, rendering it crucial for diverse applications in optical communication, sensing, and signal processing.

An interesting aspect in the development of microcavities is the composition and geometry of the distributed Bragg reflectors (DBRs). While the conventional III-V semiconductor DBRs, based, for instance, on AlGaAs/GaAs mirror pairs, have been frequently employed as mirrors in quantum light sources and microlasers, dielectric DBRs can be a promising alternative. One such example is a dielectric DBR made up of SiO\textsubscript{2}/SiN\textsubscript{x} mirror pairs, which feature a much higher refractive index contrast as compared to conventional ones, leading to higher reflectivity for a given number of mirror pairs. \cite{DielectricVCSELs, DielectricModeVolume}

In addition to the high reflectivity, dielectric DBRs based on high bandgap materials allow one to reduce absorption losses of the pump light, thereby improving the temperature stability of optically pumped microcavity lasers because of reduced heating at high excitation powers \cite{DielectricDBR}. In addition to the functionality, dielectric DBRs are fabricated using simpler, more adaptable, and less expensive deposition techniques compared to complex epitaxial growth techniques. There is no need for lattice-matched epitaxial growth, and no higher-temperature growth steps for epitaxial DBRs take place that might affect the structural properties of QDs. They offer design and fabrication flexibility and can be optimized over a wide spectral range from about 300 nm to 3 $\mu$m, making them more versatile for different quantum light sources and laser applications at different wavelengths.

In this work, we demonstrate optically pumped high-$\beta$ quantum dot microcavity lasers based on photonic-defect based hybrid microcavity structures. We numerically optimize the cavity design and systematically investigate the effects of defect size and the number of top mirror pairs on the cavity \emph{Q}-factor and the lasing properties of the microcavities. The experimental results are described in very good agreement with theoretical predictions, allowing us to obtain valuable insight into the underlying physics. Our investigations concerning the lasing behavior of the defect-based microcavities reveal excellent temperature stability and great promise for future applications, including electrically driven devices with an easily applicable intra-cavity p-type top contact.

\section{DEVICE DESIGN AND FABRICATION}

In this section, we first introduce the proposed microcavity device design and its numerical optimization in terms of \emph{Q}-factor and mode confinement. Subsequently, we present the fabrication technology to produce microcavities centered around photonic defects, utilizing upper dielectric DBR mirrors. 

\subsection{Device design}
We initiate the discussion by examining the device design requirements for quantum dot microlasers. The fundamental layer design comprises a three-layer QD active region embedded between the bottom and top DBRs, forming a planar microcavity structure. To enhance the light-matter interaction, photonic defects are introduced into the cavity containing the QD active region using wet-chemical etching, leading to lateral mode confinement. While the lower DBR comprises conventional AlGaAs/GaAs mirror pairs, the top DBR is based on dielectric layers made of SiO\textsubscript{2} and SiN\textsubscript{x}. A schematic representation of the described semiconductor-dielectric layer structure is shown in Fig. 1 (a).

 \begin{figure}[h]
\includegraphics[width=1\textwidth]{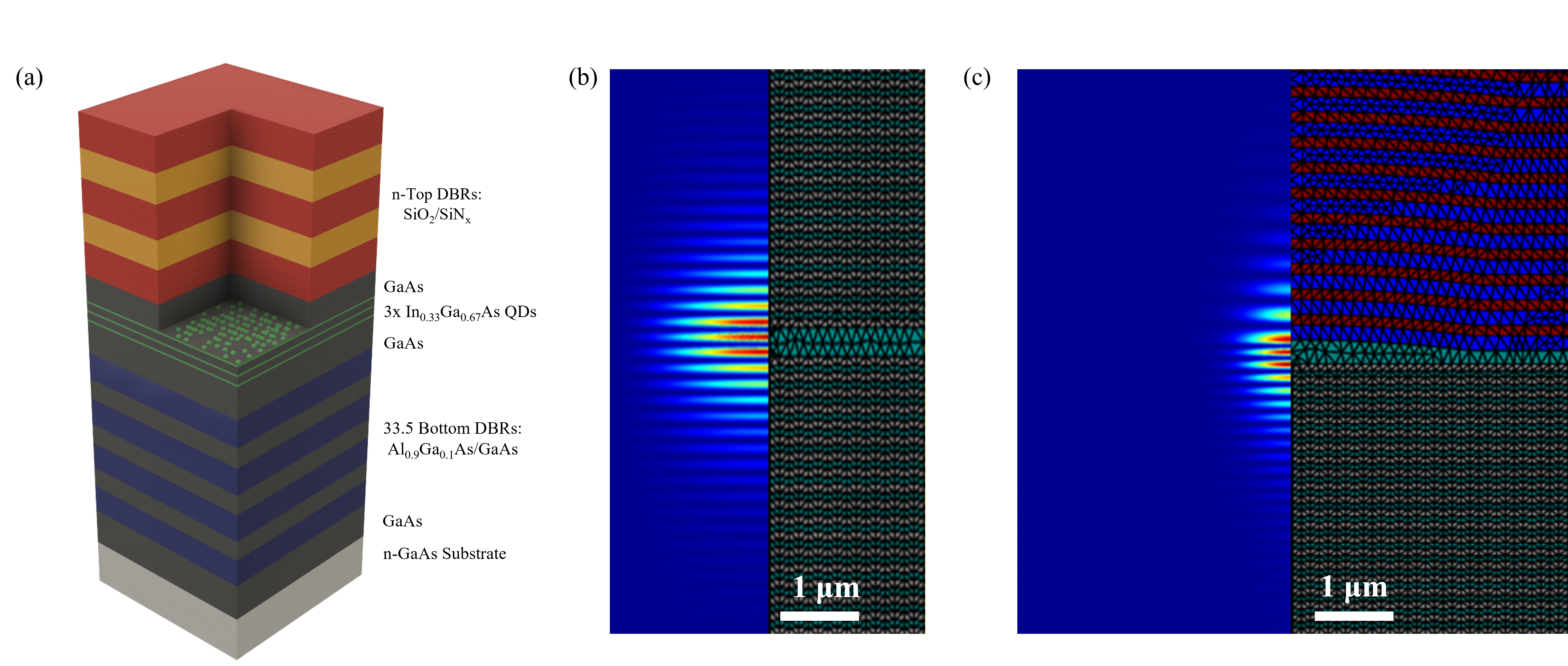}
\centering
\caption{(a) Schematic representation of the III-V epitaxial layer structure with the dielectric DBR on top. FEM simulation results of electric field strength showing the optical mode confinement for (b) a conventional micropillar design, and (c) our hybrid microresonator design incorporating a photonic defect. In the case of (c), the effective mode volume is 5 times smaller than for (b) for the same diameter of 4 $\mu$m of the integrated photonic-defect and micropillar, respectively.}
\end{figure}

The transfer matrix method (TMM) and the finite element method (FEM) are employed to optimize the design of the planar microcavity structure and the photonic defect-based microcavity devices. The numerical framework for the latter approach is provided by the software package JCMsuite from JCMwave. First, we perform TMM calculations to obtain optimal layer thicknesses of the planar microcavity. In these calculations, we consider the optical constants of the used materials like GaAs, AlGaAs, SiO\textsubscript{2}, and SiN\textsubscript{x} determined at cryogenic temperatures, which are partially obtained by extrapolation and interpolation from literature values.\cite{Refractiveindex, Luke, Gao} Via FEM simulations performed with JCMsuites's resonance mode solver with cylindrical rotational symmetry, the lateral mode confinement is studied and optimized. For the same set of shared simulation parameters, a comparison is made between conventional micropillars and our quasi-planar microresonator design incorporating photonic defects. To investigate the correspondence of the simulation and the experimental data at the end, an adjustable design parameter, analogous to the wet chemical etching parameters in later the fabrication process, is introduced in the simulations. For the simulation design of our photonic-defect structure, the general equation of a parabola, ${y = y_o + ax^2}$, is used, where $x$ and $y$ are the lateral and vertical dimensions, $y_o$ is the layer thickness, and $a$ is the coefficient which governs the width (and depth) of the photonic-defect. For a downward opening parabola, the value of $a$ is negative. The maximum value of $a$ can be 0 (planar structure) and the minimum value depends upon the layer thickness until the layers become self-intersecting ($y$<0). By precisely controlling and adjusting this coefficient, the optimal etching depth for the cavity, which in practice also changes the effective size of the photonic-defect, is simulated. The effective size of the structure in this case is determined by the reduced size of the top part of the defect as a consequence of wet-chemical etching (discussed below).
In the case of a conventional micropillar design having a 33.5x Al\textsubscript{0.9}Ga\textsubscript{0.1}As/GaAs bottom DBR and 24x Al\textsubscript{0.9}Ga\textsubscript{0.1}As/GaAs top DBR, similar to the structure as used in Ref.\cite{Kreinberg2017}, the simulated \emph{Q}-factor is approximately 68000 for a 4 $\mu$m diameter structure with a mode volume of 2 $\mu$m\textsuperscript{3} (shown in Fig. 1 (b)). For a hybrid micropillar of the same size and the bottom layer design of 33.5x Al\textsubscript{0.9}Ga\textsubscript{0.1}As/GaAs DBR, the substitution of the conventional top DBR with only 19x dielectric SiO\textsubscript{2}/SiN\textsubscript{x} mirror pairs results in an enhanced \emph{Q}-factor of 76000 and a reduced mode volume of 1.25 $\mu$m\textsuperscript{3} attributed to the improved vertical confinement due to the dielectric material. Further, by integrating our photonic-defect design with the dielectric top DBRs, we can achieve additional improvements in terms of higher \emph{Q}-factor and lower mode volume. It is also important to consider the tradeoff between \emph{Q}-factor and mode volume influenced by the cavity etching parameter (Supplementary information: Section I)\cite{ding2013vertical}. Etching a photonic-defect into the cavity reduces the effective size of the structure, which can potentially lead to increased lateral losses and thus a reduction in the \emph{Q}-factor as well, as typically observed for micropillar cavities~\cite{Reitzenstein2010}. However, it also enhances the optical confinement and results in a very low mode volume. For instance, in the specific case of 33.5x Al\textsubscript{0.9}Ga\textsubscript{0.1}As/GaAs bottom DBR, 19x top dielectric SiO\textsubscript{2}/SiN\textsubscript{x} DBR, and an etching depth parameter corresponding to 60 nm, an even higher \emph{Q}-factor of 83000 is predicted for a 4 $\mu$m parabolic photonic-defect structure with a mode volume of around 0.4 $\mu$m\textsuperscript{3}. Our numerical results yield that the hybrid defect-based microresonator structure exhibits an approximately five times smaller mode volume and a higher \emph{Q}-factor than a conventional micropillar with the same diameter. These results are crucial in understanding the combined effect of the dielectric mirror pairs along with the integration of photonic-defect into the cavity to achieve stronger vertical as well as lateral confinement within the microcavity. The simulated field distribution for our structure (shown in Fig. 1 (c)) visualizes the enhanced optical confinement in the active region as compared to the conventional micropillar design (Fig. 1 (b)).

%

\subsection{Sample growth and processing}
The fabrication process of the hybrid microresonator involves a combination of epitaxial growth, layer deposition, and processing techniques. Detailed information regarding the sample growth using metal-organic chemical vapor deposition (MOCVD) is given in the Supplementary Information (Section II).\par
 \begin{figure}[h]
\includegraphics[width=1\textwidth]{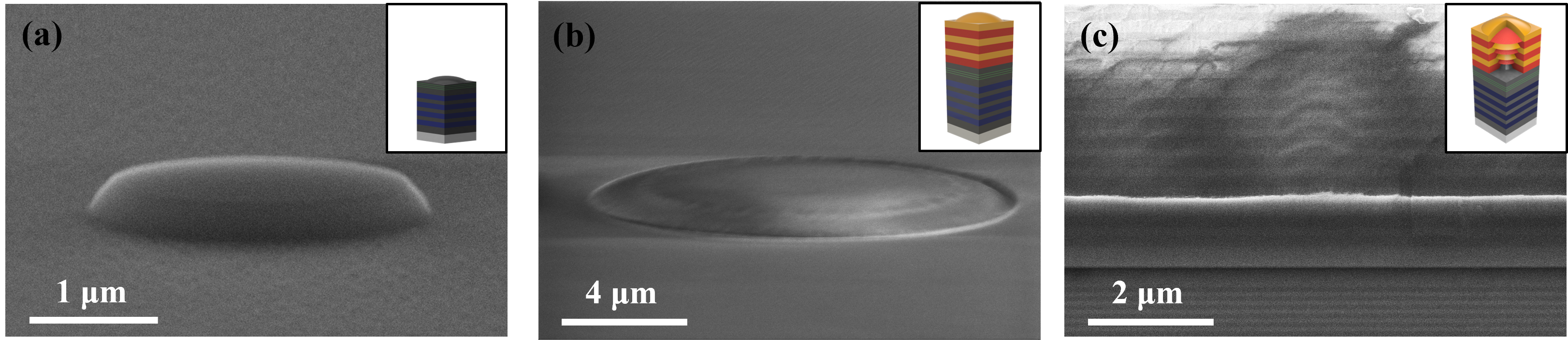}
\centering
\caption{SEM images along with associated schematic representations (insets) of (a) a central one-$\lambda$/\emph{n} GaAs cavity etched into a parabolic photonic-defect (b) a final quasi-planar microcavity sustaining the parabolic profile after the deposition of dielectric DBRs, and (c) cross-sectional view of the final structure.}
\end{figure} 

After the MOCVD growth, photonic defects are processed on the upper part of the one-$\lambda/n$ GaAs cavity. Initially, 300 nm of SiN\textsubscript{x} is deposited on the sample using plasma-enhanced chemical vapor deposition (PECVD). This layer acts as a hard mask for the subsequent etching process. The sample is then coated with a negative-tone resist, patterned using electron beam lithography (EBL), followed by resist development. In the next step, SiN\textsubscript{x} is etched away from the unexposed areas using reactive ion etching (RIE). To achieve the desired parabolic photonic-defect, the upper part of the GaAs cavity, which contains the three stacked QD layers as the active medium, undergoes wet-chemical etching. An etching solution consisting of sulphuric acid (H\textsubscript{2}SO\textsubscript{4}), water (H\textsubscript{2}O), and hydrogen peroxide (H\textsubscript{2}O\textsubscript{2}) is used. The key to optimizing the etching process is adjusting the concentration of H\textsubscript{2}O\textsubscript{2} to control the etching rate and depth, ensuring the desired parameters are achieved. In our case, the optimized concentration of H\textsubscript{2}SO\textsubscript{4}: H\textsubscript{2}O: H\textsubscript{2}O\textsubscript{2} is 1000:200:1. The final fabricated structures are etched about 100 nm to achieve a notably low mode volume while maintaining a relatively high \emph{Q}-factor (Supplementary Information: Section I). A scanning electron microscopy (SEM) image of the parabolic photonic-defect fabricated in this way is shown in Fig. 2 (a). The upper part of the structure consists of multiple PECVD-deposited dielectric SiO\textsubscript{2}/SiN\textsubscript{x} DBR layers having $\lambda$/4$n$ thicknesses. Utilizing such dielectric DBRs represents a rather straightforward and convenient fabrication approach, particularly in the context of photonic-defect cavities. Indeed, incorporating dielectric DBRs eliminates the need to reintroduce the processed sample into the MOCVD reactor. This simplified technique offers efficiency and flexibility in creating the desired structures without additional challenges associated with repeated epitaxial processes. Moreover, additional mirror pairs can be easily added or removed to optimize the device performance if necessary. Fig. 2 (b) shows the SEM image of the final structure. As we increase the number of top deposited layers, the lateral extent of the parabolic defect in the structure becomes broader, resulting in a flatter curvature and an adiabatic mode expansion. Consequently, the final structure is observed to have a significantly larger size than the defect itself. This is beneficial for light outcoupling with small beam divergence and can ensure high coupling efficiency with low numerical aperture optical fibers. The SEM image of the cross-section of the final structure in Fig. 2 (c) clearly illustrates the presence of bottom III-V DBR mirror pairs, top dielectric DBR mirror pairs, and the propagation of the parabolic photonic-defect throughout the layers of the upper DBR. It is interesting to observe that the defect remains visible even after the deposition of 19 layers of top DBR. By implementing a tailored layer design and a photonic defect within the structure, the desired optical properties can be achieved without the requirement of additional etching into a micropillar. This equally simple and attractive approach streamlines the fabrication process and minimizes optical losses that arise, for instance, from etched surfaces and other structural imperfections in the case of micropillar cavities. Moreover, it improves the temperature stability of the device due to better heat conductance. Four such quasi-planar microcavity structures with lateral mode confinement are prepared, having 5, 10, 15, and 19 top dielectric mirror pairs. Arrays of structures with defect diameters ranging from 1.5 $\mu$m to 2.5 $\mu$m are patterned on these samples.

\section{OPTICAL CHARACTERIZATION}

In this section, we optically investigate the final structures of the four samples after all device-processing steps. In order to experimentally support the theoretical and numerical results, systematic studies related to the dependence of defect diameter and number of top dielectric mirror pairs on the \emph{Q}-factor have been performed. Power-dependent input-output measurements have also been performed on the samples to determine the lasing characteristics and analyze and highlight the benefits offered by our photonic-defect cavity design based on hybrid DBRs.

\begin{figure}[h]
\includegraphics[width=1\textwidth]{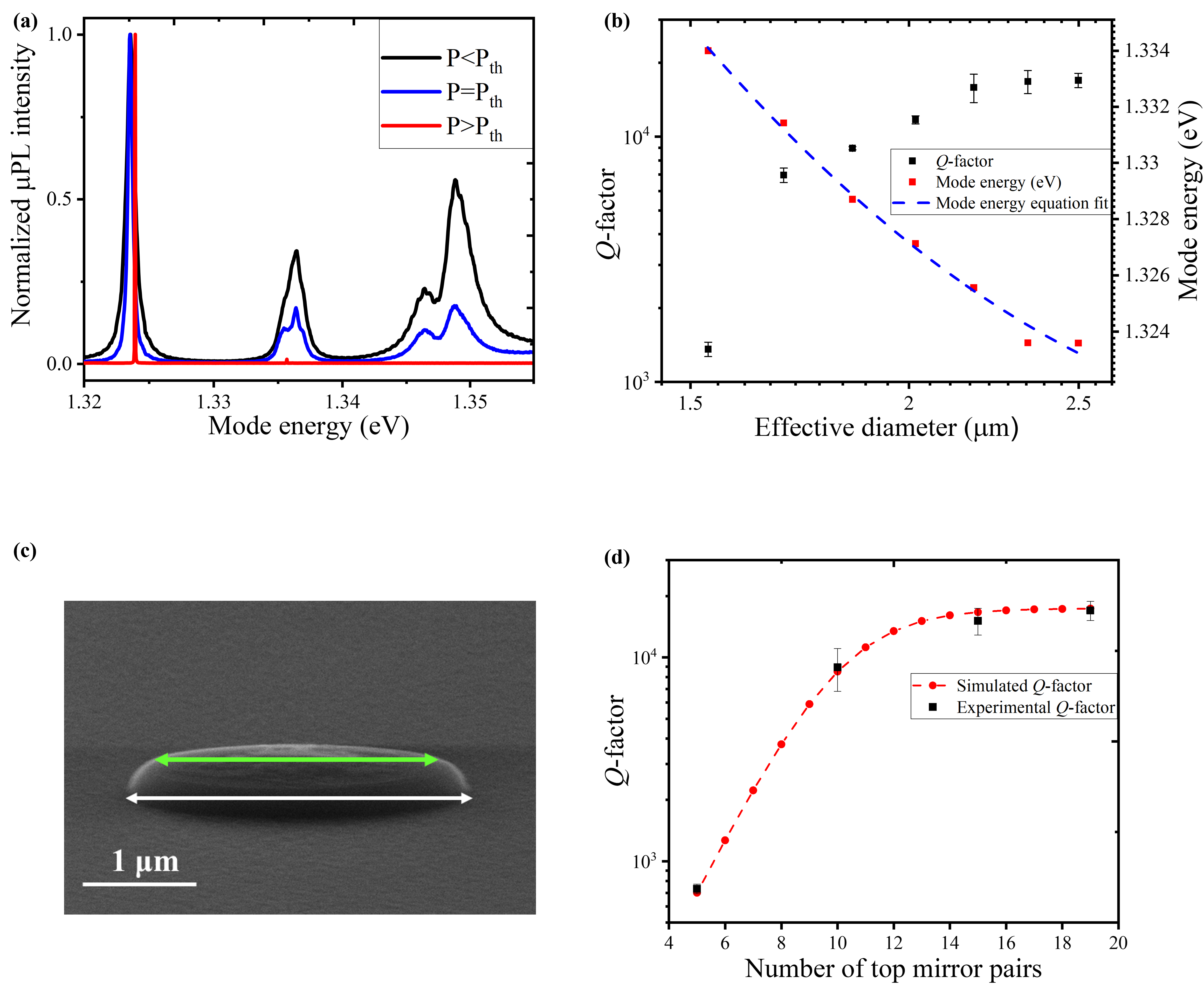}
\centering
\caption{(a) Normalized $\mu$PL spectra for excitation powers lower than, equal to, and higher than the lasing threshold. The investigated sample has 19 top DBRs and a photonic-defect cavity with an effective diameter of 2.35 $\mu$m. (b) Dependence of \emph{Q}-factor and mode energy on the diameter of the integrated photonic-defect. (c) Photonic-defect structure showing an overall reduction in size due to wet-chemical etching (depicted with green and white arrows). The green arrow shows the effective diameter of 2.35 $\mu$m of the photonic-defect cavity. (d) Dependence of the \emph{Q}-factor on the number of mirror pairs in the top dielectric DBR.}
\end{figure} 

The four samples are investigated using low-temperature microphotoluminescence ($\mu$PL) spectroscopy using the experimental setup described in the Supplementary Information (Section III). In Fig 3 (a), the normalized emission spectra are shown for the sample having 19 top DBRs and a photonic-defect cavity with an effective diameter of 2.35 $\mu$m. These exemplary spectra correspond to three different pump powers: lower than, equal to, and higher than the lasing threshold. It is evident that when the input power exceeds the lasing threshold, a narrow emission peak corresponding to the fundamental cavity mode (HE\textsubscript{11}) is observed. Conversely, for input powers less than the lasing threshold, significant higher-order modes like HE\textsubscript{21}, HE\textsubscript{12}, etc. appear in the spectra. These results are similar to the case of a conventional micropillar cavity \cite{Reitzenstein2010}. Noteworthy, in our case the separation in mode energy between the fundamental mode and the first higher-order mode is approximately 12.9 meV which is considerably higher than 8 meV \cite{Reithmaier}, typically observed for micropillar cavities of similar diameter ($\sim$ 2.5 $\mu$m) as our integrated photonic defect. This result is consistent with the strongly reduced mode volume of our structures compared to micropillars with the same diameter, facilitating enhanced mode confinement in combination with low optical losses, which could even lead to single-QD strong coupling effects in future studies. 

 A comprehensive study involving systematic $\mu$PL measurements is done on an array of photonic-defect cavity structures with varying effective diameters between 1.5 $\mu$m and 2.5 $\mu$m in steps of approximately 160 nm, and \emph{Q}-factors near transparency are measured. In Fig. 3 (b), the \emph{Q}-factor for the smallest structure is approximately 1400 and the largest structure is approximately 17000. The observed behavior of the mode energy shows the expected blueshift of about 10 meV in the studied range with decreasing diameter due to increased lateral mode confinement. To model the experimentally observed trend, a modified version of the well-known diameter-dependent mode energy relation of micropillar cavities is used \cite{ModeEnergyGutbrod}
 \[E_c = \sqrt{E_o^2+\frac{\hbar^2c^2}{\varepsilon_r}\frac{x^2_{n_\varphi,n_r}}{R_{eff}^2}}\]
where $E_c$ is the resonance energy of the cavity mode, $E_o$ is the resonance energy of the planar cavity, $R_{eff}$ is the effective radius of the structure, $x_{n_\varphi,n_r}$ is the $n_r^{th}$ zero of the Bessel function $J_{n_\varphi}(x_{n_\varphi,n_r}/R_{eff})$, and $\varepsilon_r$ is the effective dielectric constant of the material of the cavity. The experimentally determined diameter-dependent mode energies closely align with the calculated values derived from this theoretical formula. The effective radius, $R_{eff}$, as employed in the formula, is quantified through measurements derived from SEM images of the photonic-defect cavity. In practice, as the etching rate or time is increased, there is a progressive reduction in the size of the top part of the photonic-defect structure (effective size) as shown in Fig. 3 (c). The effective size of the photonic-defect structure can be easily adjusted during the device processing by adapting the wet-chemical etching parameters, such as etching depth and rate.\par

Fig. 3 (d) shows the dependence of the \emph{Q}-factor on the number of top dielectric mirror pairs measured near transparency. For this study, structures with an effective diameter of 2.35 $\mu$m on four different samples with varying numbers of mirror pairs in the top dielectric DBR, are investigated. As anticipated, while the sample with only 5 mirror pairs in the top DBR shows a clear lateral mode pattern, the \emph{Q}-factor is found to be relatively low ($\sim700$). As the number of layers increases, the \emph{Q}-factor shows a consistent upward trend before reaching saturation at about 15 mirror pairs. It is observed that 10 mirror pairs in the top DBR are enough to achieve lasing. However, a significantly lower lasing threshold can be achieved by increasing the number of mirror pairs in the upper DBR to 15 and, eventually, to 19, due to reduced optical losses (see discussion below). For the same layer and cavity design parameters, the simulated values of \emph{Q}-factor obtained from FEM simulations show excellent agreement with the experimental \emph{Q}-factors. 

\begin{figure}[h]
\includegraphics[width=0.65\textwidth]{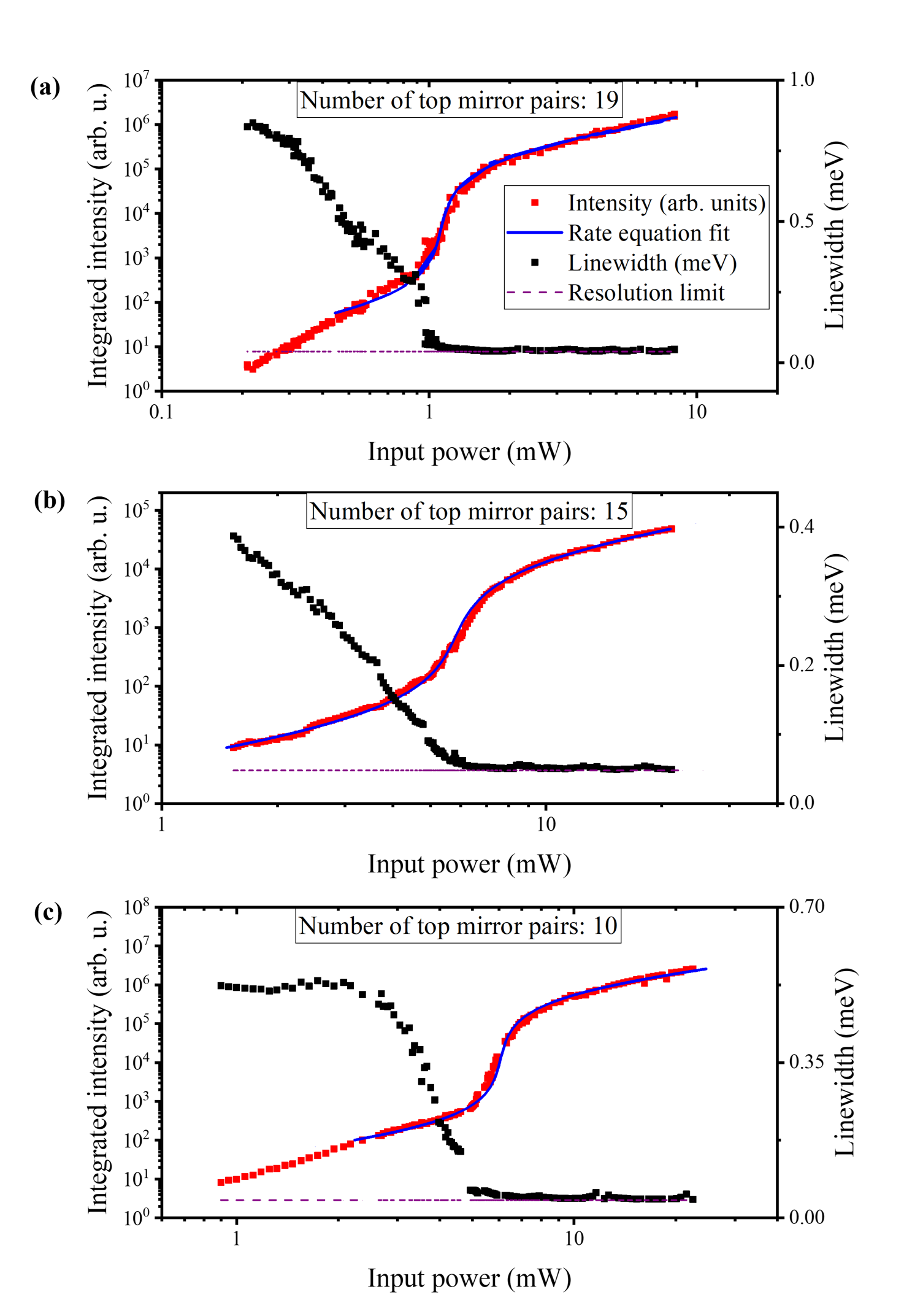}
\centering
\caption{(a) Pump power-dependent input-output characteristics and spectral linewidth for (a) 19, (b) 15, and (c) 10 mirror pairs in the top dielectric DBR.}
\end{figure} 
Fig. 4 (a), (b), and (c) exhibit the power-dependent input-output characteristics on a double logarithmic scale for a structure with an effective diameter of 2.35 $\mu$m having 19, 15, and 10 mirror pairs, respectively in the top DBR. The intensities have been determined by fitting the $\mu$PL emission spectra with Pseudo-Voigt line-shape fitting\cite{ArisLineshape}. These plots show a distinct non-linear S-curve along with a significant reduction in linewidth, serving as a clear indication of lasing behavior. It is important to highlight that at higher powers, the linewidth becomes constrained by the resolution limit (40 $\mu$eV) of the spectrometer used in the experiment. The blue line in the plots depicts the laser rate equation fitting using the same model as applied in Ref.\cite{YamamatoLaserRate}. In our case of rate-equation fitting, the exciton number at the transparency threshold, $n_o$ is set to 9000 aligning with the gain due to three stacked layers of QDs.\cite{Andreoli} This yields a $\beta$-factor equal to 1.5*10\textsuperscript{-2} (1.5\%) lasing threshold of approximately 900 $\mu$W for the case of 19 top dielectric mirror pairs. The comparatively high-$\beta$ factor and low lasing threshold power are noteworthy given that the final structure has not undergone etching, and the entire lateral mode confinement is solely achieved due to the presence of the introduced photonic defect within the cavity. In Fig. 4 (b) and (c), the lasing thresholds are measured to be around 3.5 mW and 5 mW, with $\beta$-factors equal to 1.6*10\textsuperscript{-3} (0.16\%) and 2*10\textsuperscript{-4} (0.02\%) for 15 and 10 mirror pairs in the top DBR, respectively. The systematic increase (decrease) of the threshold pump power ($\beta$-factor) with decreasing number of mirror pairs in the upper DBR is consistent with the associated reduction of the \emph{Q}-factor. Our results are very promising and provide evidence for the advantages of the photonic-defect-based cavity design concept in combination with dielectric DBRs, which enables lasing in structures without laterally etched sidewalls and a reduced number of mirror pairs in the top DBR. Additionally, our microlaser concept shows excellent temperature stability, avoiding temperature-induced redshift and spectral broadening at high excitation powers typically observed for conventional nano- and microlasers\cite{HeatingBGayral, HeatingeffectsJaffrennou, Nakwaski1997, Li2017} (discussed in Supplementary Information - Section IV).

\section{CONCLUSION}
In this work, we have realized and studied QD microcavity lasers by leveraging photonic-defect cavity engineering combined with the incorporation of dielectric mirrors. The integration of photonic-defects facilitated low mode volume and enhanced optical confinement in a quasi-planar resonator geometry. By exploiting the material properties of dielectric DBR and optimized fabrication characteristics, we achieved significant advancements in resonator \emph{Q}-factors, laser performance, and device design. The methodical investigation of the effect of the number of mirror pairs in the top DBR on \emph{Q}-factor closely aligns with the simulation results, affirming that even as few as 10 mirror pairs are sufficient to attain lasing in such structures. Particularly, our quasi-planar microresonator comprising 19 top dielectric mirror pairs demonstrated a high $\beta$-factor of 1.5\%, without an additional need of laterally etching the final structure into a micropillar. The successful demonstration of lasing behavior in structures with 15 and 10 dielectric DBR further proves the merit and flexibility of our innovative design and material implementation. The devices also exhibited notable resilience against heating effects at higher pump powers, demonstrating their suitability for effective device operation. It is important to note that such advancements hold great potential not only for lasing applications but also for the realization of single-photon sources. Further technological advances in terms of cavity engineering and fabrication techniques can pave the way for impactful contributions to both photonics and quantum technologies.

\begin{acknowledgments}
 This work was funded by the German Research Foundation (RE2974/20-1, RE2974/33-1, and INST 131/795-1 320 FUGG), and by the German Federal Ministry of Education
and Research (BMBF) through the Project MultiCoreSPS (Grant No. 16KIS1819K). The authors would like to thank Kathrin Schatke and Praphat Sonka for their expert technical support.   
\end{acknowledgments}


\end{document}


\author{Kartik Gaur}
 \author{Ching-Wen Shih}
 \author{Imad Limame}
 \author{Aris Koulas-Simos}
 \author{Niels Heermeier}
 \author{Chirag C. Palekar}
 \author{Sarthak Tripathi}
 \author{Sven Rodt}
 \author{Stephan Reitzenstein}
 \email[Corresponding author: ]{stephan.reitzenstein@physik.tu-berlin.de}
\affiliation{Institut für Festkörperphysik, Technische Universität Berlin, Hardenbergstraße 36, D-10623 Berlin, Germany
}%

\title{High-$\beta$ lasing in photonic-defect semiconductor-dielectric hybrid microresonators with embedded InGaAs quantum dots: Supplementary Information} 
\pacs{}
\maketitle 

\section{Simulation results}

\subsection{Comparison between conventional micropillar, hybrid micropillar, and hybrid photonic-defect based microresonator:}
For a 2.35 $\mu$m structure having 33.5x Al\textsubscript{0.9}Ga\textsubscript{0.1}As/GaAs bottom DBR, we conduct a comparative simulation analysis involving three different microcavity configurations: a conventional micropillar with 19x Al\textsubscript{0.9}Ga\textsubscript{0.1}As/GaAs top DBR, a semiconductor-dielectric hybrid micropillar with 19x top dielectric SiO\textsubscript{2}/SiN\textsubscript{x} DBR, and a photonic-defect incorporated hybrid microcavity structure with 19x top dielectric SiO\textsubscript{2}/SiN\textsubscript{x} DBR and a photonic defect etching depth of 100 nm. The conventional micropillar shows a \emph{Q}-factor of 15000 and a mode volume of 0.6 $\mu$m\textsuperscript{3}. Substituting the top semiconductor DBR with dielectric DBR results in a similar \emph{Q}-factor of around 15000, but the mode volume gets reduced to 0.45 $\mu$m\textsuperscript{3} attributed to the enhanced vertical confinement due to the dielectric material. However, when the photonic defect of the same size gets integrated into the cavity of the hybrid layer structure, a significant increase in the \emph{Q}-factor to 17000 is observed. Simultaneously, the mode volume is remarkably reduced to 0.28 $\mu$m\textsuperscript{3}. This notable enhancement in the \emph{Q}-factor and reduction in the mode volume can be associated with a combination of factors including the favorable material properties (vertical confinement) of the dielectric DBR and the enhanced lateral confinement facilitated by the photonic defect. Fig S 1. shows the FEM simulations for the three cases.
\begin{figure}[h]
\includegraphics[width=0.9\textwidth]{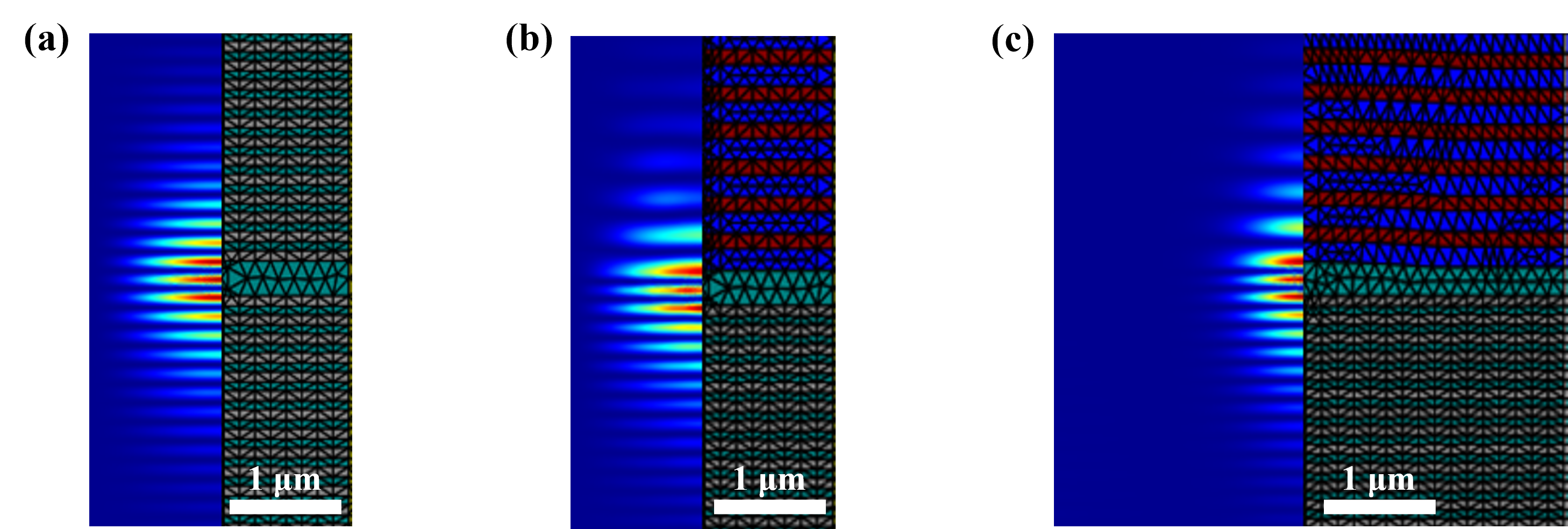}
\centering
\caption{FEM simulation results of electric field strength showing the optical mode confinement for (a) a conventional micropillar design, (b) a hybrid micropillar design, and (c) hybrid microresonator design incorporating a photonic defect.}
\end{figure}

\subsection{Effect of cavity etching depth on \emph{Q}-factor and mode volume:}
In structures featuring an incorporated photonic defect, the vertical dimension of such parabolic photonic-defect, determined (and adjusted) by the parameters of wet-chemical etching, has an important role in determining the mode volume and \emph{Q}-factor of the microcavity. As the etching depth increases, the curvature of this parabolic defect also increases, resulting in an overall reduction in the effective size of the structure. Consequently, the \emph{Q}-factors show a notable reduction. However, the mode volume of these structures is significantly reduced. Understanding this tradeoff between \emph{Q}-factor and mode volume is of utmost importance for various device applications. In some applications (e.g. laser), a high \emph{Q}-factor is necessary and can be achieved by introducing a photonic defect with minimal etching depth. Conversely, for applications such as light coupling into optical fibers (single-mode fibers in telecom applications), deeper etching depths can yield substantially smaller mode volumes, a critical requirement for such purposes. In our microcavity design, we selected an etching depth of 100 nm. This depth strikes a balance, providing a high enough \emph{Q}-factor for lasing application while simultaneously offering the potential to employ these structures in fiber-coupled devices for future applications.

\begin{figure}[h]
\includegraphics[width=0.75\textwidth]{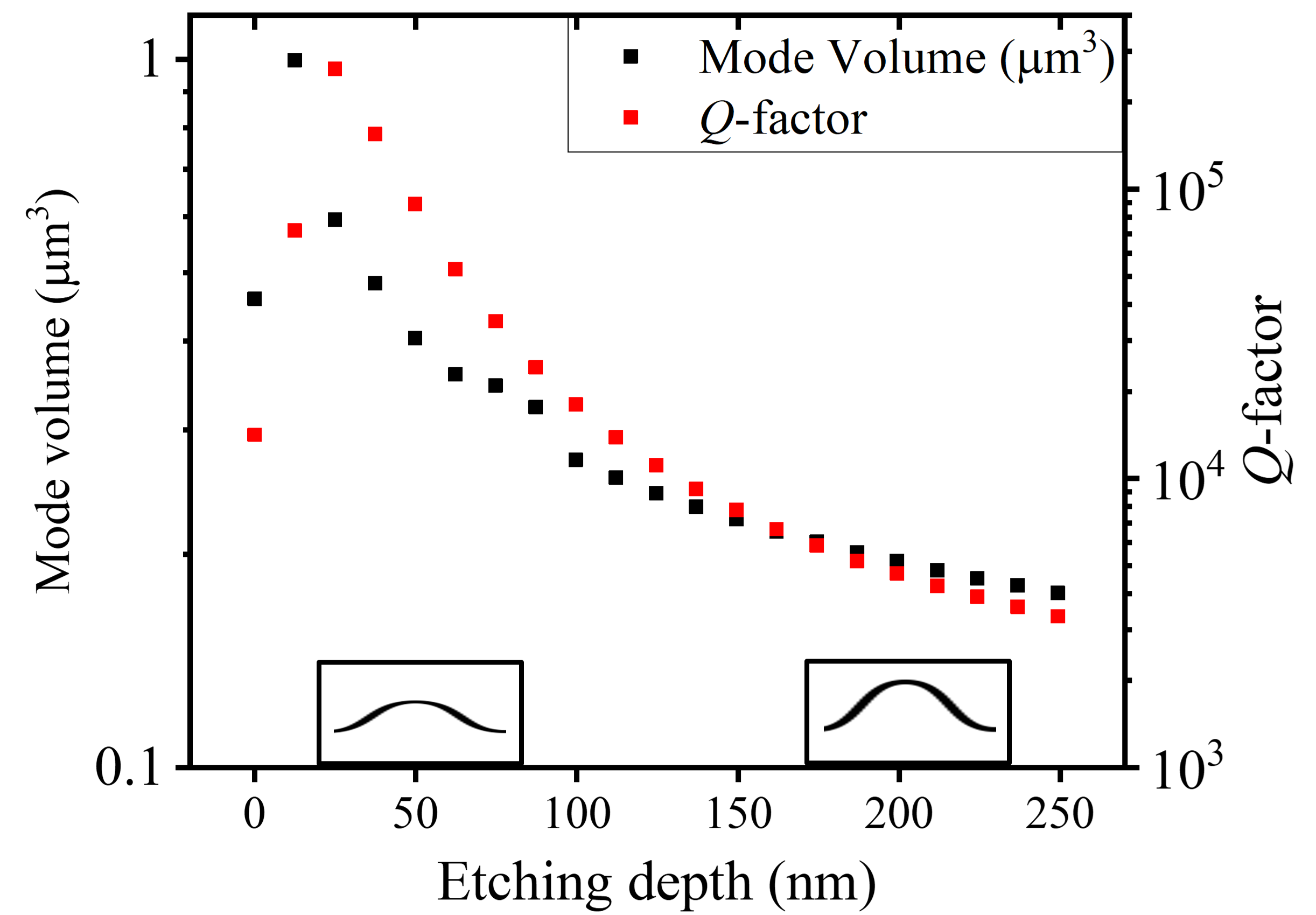}
\centering
\caption{Variation of mode volume and \emph{Q}-factor with etching depth (inset: schematic representation)}
\end{figure}

\section{Sample growth}
Epitaxial layers are grown using metal-organic chemical vapor deposition (MOCVD), employing metalorganic precursors such as TMGa, TMAl, TMIn, AsH\textsubscript{3}, and TBA, with H\textsubscript{2} as the carrier gas. The InGaAs quantum dots are grown at an optimized temperature of 500 \degree C. The bottom DBRs and the spacer layers between the QDs are grown at temperatures of 700 \degree C and 615 \degree C, respectively. The bottom part of the structure consists of a 300 nm GaAs buffer layer grown on a Si-doped GaAs (100) wafer. This is followed by  DBR consisting of 33.5 $\lambda$/4$n$-thick layer pairs of Al\textsubscript{0.9}Ga\textsubscript{0.1}As/GaAs. Three layers of In\textsubscript{0.33}Ga\textsubscript{0.67}As are grown for the formation of stacked QDs, with each QD layer being separated by 19 nm thick GaAs spacer layers. These QD layers are incorporated within the central one-$\lambda$/\emph{n} thick GaAs cavity with a high density of around $10^{10}$/cm\textsuperscript{2}, ensuring high optical gain.

\section{Experimental setup}
The experimental measurements are conducted utilizing a helium flow cryostat operating at a temperature of 20 K, with a 781 nm laser as the excitation source. A variable magnification beam expander is incorporated in front of the laser to precisely control the laser beam size. This adjustment allows for achieving a focused spot size of around 2 $\mu$m. Furthermore, an optical objective with a numerical aperture (NA) of 0.4 is placed before the cryostat. To enable precise spatial adjustment of the microscope objective, an extra $x/y/z$ piezo stage is incorporated, ensuring fine positioning of the objective with respect to the structures. The collected signal from these microresonators is captured using the same microscope objective. Subsequently, a grating spectrometer is employed for detection, offering a spectral resolution of approximately 40 $\mu$eV. To selectively detect $\mu$PL signals from individual structures, a pinhole is positioned within a confocal microscope configuration. 

\section{Excitation power dependent emission wavelength}

Our photonic-defect based microlasers show excellent temperature stability. Indeed, as exemplarily shown in Fig. S 3 for the microlaser with 19 mirror pairs in the upper dielectric DBR, only a blue shift in emission is observed with increasing pump powers above threshold. In this pump regime, the free carrier density in the absorbing layers increases, which results in a carrier-induced reduction of the refractive index resulting in a blueshift in emission mode energy.\cite{Bennet1990} The absence of a temperature-induced redshift and linewidth broadening, typically observed for instance for conventional micropillar lasers \cite{Mohideen1994}, is explained by a significantly better heat conductance of the quasi-planar device design in combination with the absence of pump light absorption in the upper dielectric DBR\cite{C-WShih}. This interesting feature can ensure frequency-stable laser emission under strong excitation and is highly desirable in device operation, for instance in quantum photonics applications, when resonantly exciting single quantum emitters~\cite{Kreinberg2018}.

\begin{figure}
\includegraphics[width=0.8\textwidth]{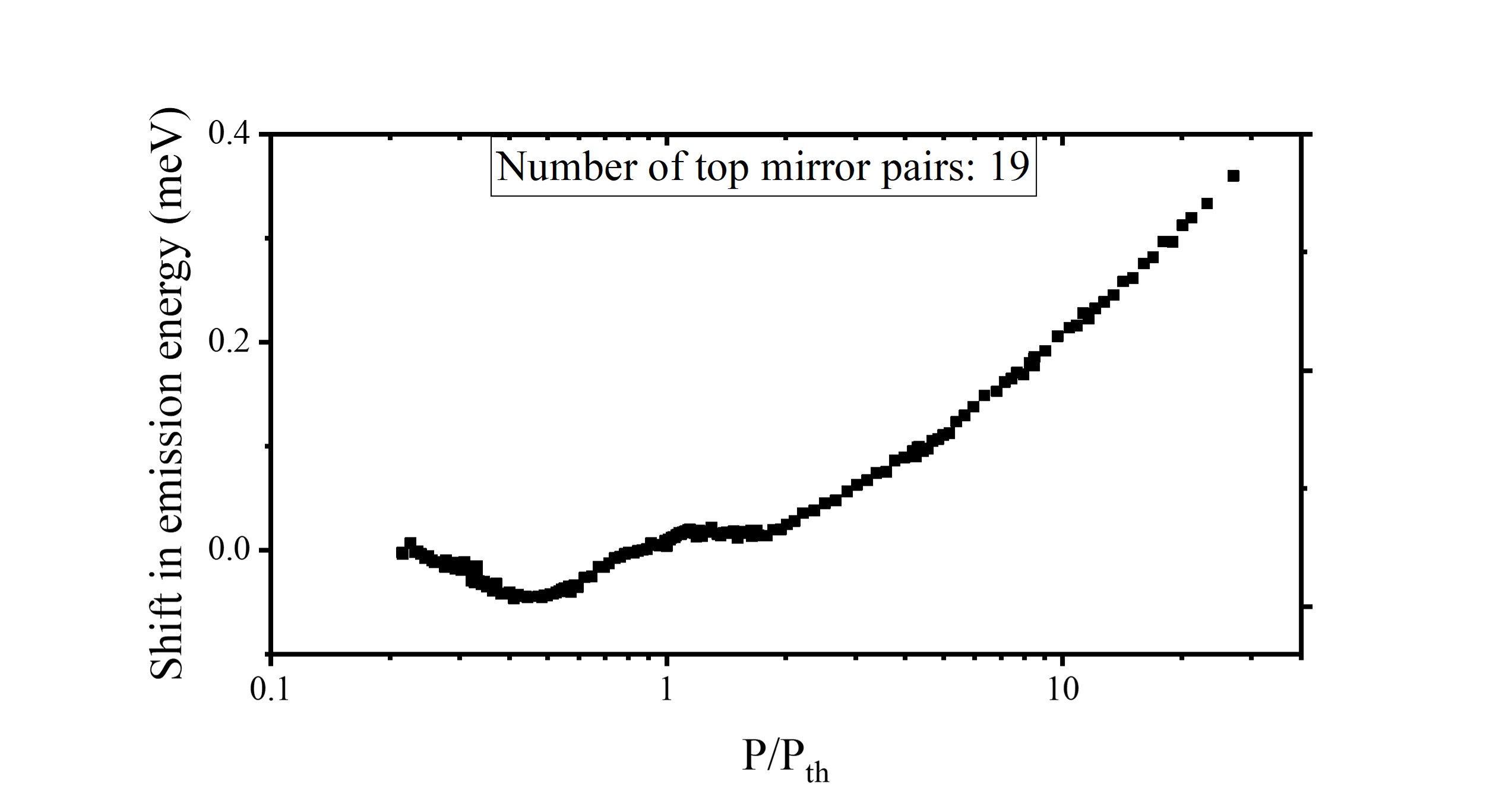}
\centering
\caption{Power-dependent mode energy characteristics corresponding to the structure with 19 mirror pairs in the top dielectric DBR, showing a significant blue shift in emission with high pump powers.}
\end{figure}
